\documentclass[superscriptaddress,pra,reprint,floatfix]{revtex4-1}
\usepackage[utf8]{inputenc}
\usepackage{lipsum}
\usepackage{braket}
\usepackage{amsmath}
\usepackage{amssymb}
\usepackage{upgreek}
\usepackage{graphicx}
\usepackage{placeins}
\usepackage{bbold}
\usepackage{dsfont}
\usepackage{hyperref}
\usepackage[dvipsnames]{xcolor}
\hypersetup{colorlinks=true, citecolor=blue, urlcolor=blue, linkcolor=blue}
\usepackage{soul}

\definecolor{lightteal}{RGB}{170,237,229}

\definecolor{lightblue}{RGB}{173,216,250}

\DeclareMathOperator{\sech}{sech}
\newcommand*\diff{\mathop{}\!\mathrm{d}}

\newcommand{\ku}{\ket{\uparrow}}
\newcommand{\kd}{\ket{\downarrow}}

\usepackage[shortlabels]{enumitem}

\makeatletter
\let\latexlabel\ltx@label
\makeatother

\begin{document}
\bibliographystyle{apsrev4-1}
\title{Impulsive Spin-Motion Entanglement for Fast Quantum Computation and Sensing}
\author{Randall Putnam}
\email{ranpjr@gmail.com}
\author{Adam D. West}
\thanks{Current affiliation: IonQ, Inc., College Park, MD 20740}
\author{Wesley C. Campbell}
\author{Paul Hamilton}
\email{ph@physics.ucla.edu}
\affiliation{UCLA, Department of Physics and Astronomy, 475 Portola Plaza, Los Angeles, California 90095, USA}
\date{\today}

\begin{abstract}
We perform entanglement of spin and motional degrees of freedom of a single, ground-state trapped ion through the application of a $16~{\rm ps}$ laser pulse. The duration of the interaction is significantly shorter than both the motional timescale ($30~\upmu\mbox{s}$) and spin precession timescale ($1~\mbox{ns}$) , demonstrating that neither sets a fundamental speed limit on this operation for quantum information processing. Entanglement is demonstrated through the collapse and revival of spin coherence as the spin components of the wavefunction separate and recombine in phase space. We infer the fidelity of these single qubit operations to be $(97^{+3}_{-4})\%$.
\end{abstract}
\maketitle

\section{Introduction}
\label{sec:introduction}
Spin-motion entanglement is at the heart of many trapped-ion quantum computers. Entanglement between the internal qubit states of ions is produced via ion-ion interactions.  These interactions are mediated by motion within the trap and modulated by the application of spin-dependent forces \cite{Cirac1995,Molmer1999,Sorensen1999,Sorensen2000,BeGates,ResilientGates}. In order to avoid the problem of spectral crowding, gates have been operated in the strong excitation regime, where the applied spin-dependent forces are impulsive, or applied much faster than the ions' motional mode period \cite{Garcia-Ripoll2003,Duan2004,Bentley2013,Steane2014}. These impulsive forces, known as spin-dependent kicks (SDKs), dynamically impart momentum to the ion, with the direction of the kick dependent upon the ion’s internal qubit state.

Previous work has demonstrated both single- and two-qubit gates with ultrafast pulses \cite{Madsen2006,Campbell2010,Mizrahi2013,Wong-Campos2017, LMDuan}. While the picosecond duration of a single pulse from a mode-locked laser makes it attractive for building gates in the strong excitation regime, single pulses do not tend to produce the desired outcome with hyperfine qubits. Single pulse operations have been performed using resonant excitation as well as stimulated Raman transitions. In the resonant case, a $\pi$-rotation was performed using a single ultrafast pulse with 98.1\% fidelity \cite{LMDuan}, but the scheme could not be used to perform arbitrary single qubit rotations.  Single-pulse, single qubit gates using stimulated Raman transitions in the hyperfine qubit of $^{171}{\rm Yb}^+$ were limited by the finite qubit splitting while two-qubit gate fidelity using single-pulse spin-dependent kicks (SDKs) was limited by multi-photon transitions that produce unwanted higher-order momentum modes \cite{Campbell2010,Mizrahi2013,Wong-Campos2017}. In both schemes, to achieve high-fidelity two-qubit gates, multi-pulse sequences that are many times longer than the single pulse duration are necessary. This in turn makes two-qubit gates longer than the attractively-short duration of the atom-light interaction in a single laser pulse.

Aside from applications in quantum information processing, high-fidelity spin-dependent kicks are also a key feature of atom interferometry. Increasing wavepacket separation by large momentum transfer beamsplitter operations enhances interferometer sensitivity \cite{Cronin2009,Cadoret2008,Muller2009,Chiow2011,Parker2018,Jaffe2018,Pagel2019}. The ability to perform high-fidelity ultrafast spin-dependent kicks would enable even higher momentum transfer, a key ingredient to the recently proposed ion gyroscope interferometer \cite{Campbell2017,West2019}.

Here we demonstrate high-fidelity, ultrafast qubit rotations and spin-motion entanglement using a single, $16\,\text{ps}$ laser pulse to drive a stimulated Raman transition in the ground-state Zeeman qubit of $^{138}$Ba$^+$. By observing the decay and revival of interference fringe visibility using a Ramsey pulse sequence, we verify the generation of spin-motion entanglement of a Zeeman qubit using a single laser pulse. Working with a Zeeman qubit offers improvements in both gate speed and simplicity compared to hyperfine qubits. The smaller qubit splitting allows for single-pulse single qubit gate fidelity comparable with current state of the art. Further, using polarization selectivity, a spin-dependent kick can be performed in a Zeeman qubit with a single laser pulse without producing higher-order momentum modes. Zeeman qubits are also the natural choice for a gyroscope interferometer, as the magnetic moment associated with the ion's motion, which can mask the desired rotation phase in the presence of a magnetic field, can be canceled by the Zeeman qubit's spin magnetic moment, essentially making this a clock qubit for the interferometer \cite{Campbell2017}.

\label{sec:experimental_setup}
\begin{figure}[!t]
    \centering    
    \includegraphics[width=1\linewidth]{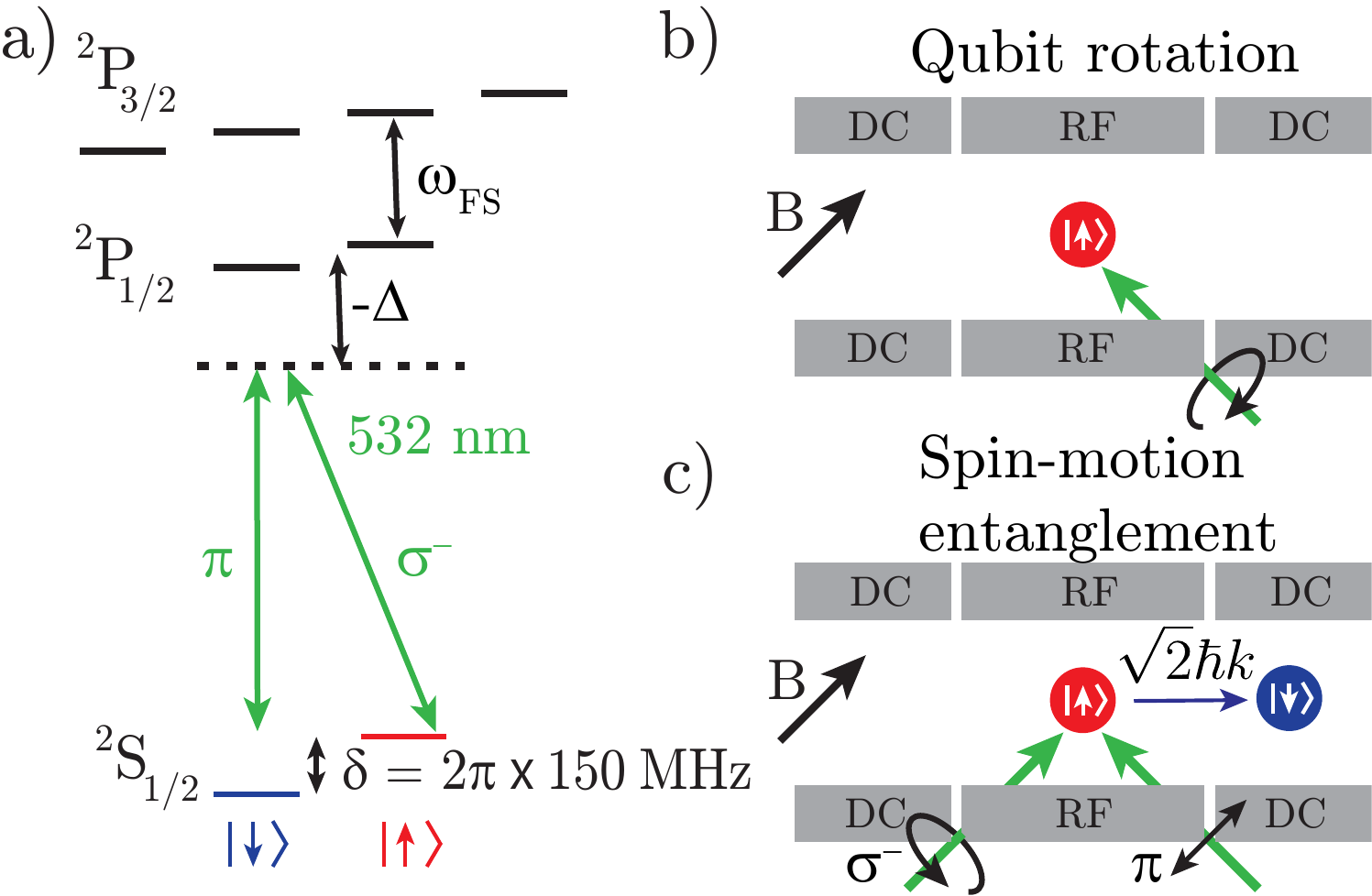}
    \caption{a) Energy levels relevant for the stimulated Raman transition. On the right, a top-down view of the ion (colored circle) with surrounding trap electrodes. The Raman beam directions with their polarizations are shown in green and the magnetic field direction in black. b) Schematic of ultrafast qubit rotation. A single, circularly-polarized beam drives the ion from $\ku$ to $\kd$. c) Schematic of spin-dependent kick, with the resulting momentum kick.}
    \label{fig:ultrafast_paper_exp_figure}
\end{figure}
\section{Experimental Setup}
Our apparatus utilizes a single ${}^{138}\mathrm{Ba}^+$ ion trapped in a four-rod, linear Paul trap with axial secular frequency $\omega=2\pi\times32.4$~kHz and radial secular frequencies $\omega_\mathrm{rad}\approx2\pi\times 100$~kHz. The stimulated Raman transition as shown in Fig.~\ref{fig:ultrafast_paper_exp_figure}(a), can be driven using pulses from a 532 nm, mode-locked Nd:YVO$_4$\footnote{Coherent Paladin SCAN 532-36000.} laser with a repetition rate of 76 MHz \cite{West_2021}. An intensity autocorrelation measurement yields a $\sech$ pulse shape, with a full width at half max of $\tau_{\rm{pulse}}=\ $16.4(5) ps, corresponding to a spectral bandwidth $\Delta f\approx 0.315/\tau_{\rm pulse}\approx19$~GHz.  Arbitrarily gated patterns of laser pulses can be generated through the use of an electro-optic pulse picker. To achieve a $\pi$ pulse within a single laser pulse, we tightly focus a beam to a 1/$e^2$ intensity radius of $w_0 =8.5(4)\,\upmu$m. Single-pulse SDKs, which require differing wavevectors, are achieved by splitting the laser beam into an additional, orthogonally directed beam with a waist of $w_0 = 20(2)\,\upmu$m. Temporal overlap of the Raman beams is achieved with an optical interferometer to measure the electric field autocorrelation and is subsequently refined using the response of the ion.
\section{Single-Pulse Single Qubit Rotations}
The finite bandwidth of the Raman pulse can only fully transfer population between degenerate levels, and any energy splitting leads to an effective detuning of the two-photon resonance, resulting in incomplete population transfer. The fidelity of a hyperbolic secant pulse with pulse area $\theta=\int \mathrm{d}t'\ \Omega(t')$ and temporal width $\tau_{\rm pulse}$ driving population between a pair of states split by $\delta$ has been provided by Rosen and Zener \cite{Rosen1932}:
\begin{align}
    \mathcal{F} &= \sin^2\left(\frac{\theta}{2}\right){\rm sech}^2(\delta\tau_{\rm pulse}/1.76).
    \label{eq:RosenZener}
\end{align}

\begin{figure}[!t]
    \centering
    \includegraphics[width=0.9\linewidth]{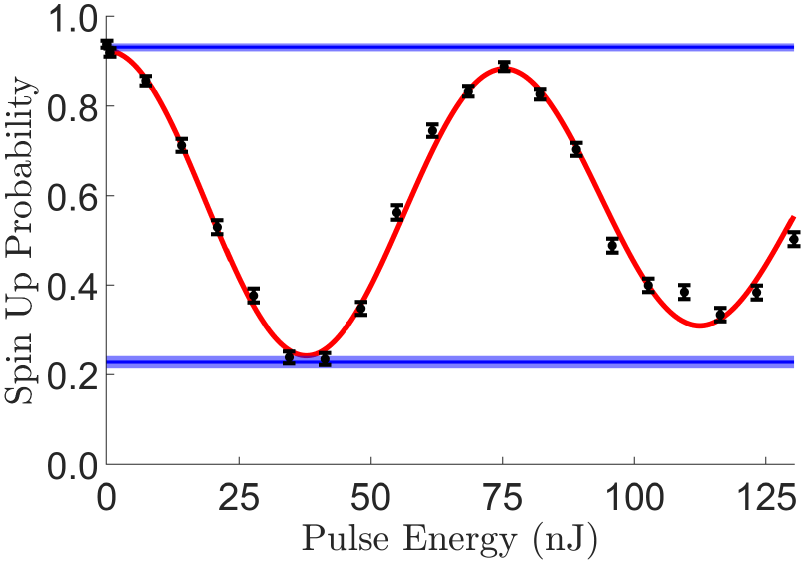}
    \caption{ Ultrafast qubit rotations: a Rabi flopping curve produced by applying a single laser pulse of varying energy to an ion initially prepared in the $\ku$ state. Black: data points with statistical error bars from 1000 repetitions of the experiment. Red: fit to the data, $\mathcal{V}_{\rm fit}=0.68$. Blue: bars indicate average state preparation and measurement (SPAM) limits for $\kd$ and $\ku$, $\mathcal{V}_{\rm SPAM}=0.70$. The width of the bars indicates the standard deviation of multiple SPAM measurements.}
    \label{fig:ultrafast_rabi_single_beam}
\end{figure}

First we demonstrate ultrafast single qubit rotations using a single, circularly-polarized beam directed orthogonal to the applied magnetic field (see Fig.~\ref{fig:ultrafast_paper_exp_figure}(b)). For the given geometry, this polarization maximizes the two-photon Rabi frequency, $\Omega_{2\gamma}$, 
\begin{align}
    \Omega_{2\gamma}&=\frac{\sqrt{2}}{6\Delta}\frac{d^2}{\hbar^2}\left(\frac{2\Delta}{\Delta-\omega_{\rm FS}}-1\right)\left(E_\pi^*E_{\sigma_-}+E_{\sigma_+}^*E_{\pi}\right),
    \label{eq:rabifreq}
\end{align}
while canceling the differential light shift,~$\delta_{2\gamma}$,
\begin{align}
        \delta_{2\gamma}&=\frac{d^2}{6\hbar^2\Delta}\left(\frac{4\Delta}{\Delta-\omega_{\rm FS}}+1\right)\left( |E_{\sigma^+}|^2-|E_{\sigma^-}|^2\right),
        \label{eq:lightshift}
\end{align}
with $\omega_{\rm FS}$ the fine structure splitting, as shown in Fig.~\ref{fig:ultrafast_paper_exp_figure}(a); $\Delta=\omega-\omega_0$ the difference between the laser frequency $\omega=c|\vec{k}|$, with $\vec{k}$ the laser wavevector, and the $^2\mathrm{S}_{1/2}\rightarrow~^2\mathrm{P}^o_{1/2}$ transition resonance, $\omega_0$; $d$ the dipole moment of the same transition; and $E_j$ the complex electric field amplitude for polarization $j$. The beam waist is measured to be $w_0= 8.5(4)~\upmu$m and the maximum pulse energy is 129(5)~nJ. Preparing the ion in $\ku$, applying a single laser pulse with varying energy, and reading out the final state we map out Rabi flopping curves such as that shown in Fig.~\ref{fig:ultrafast_rabi_single_beam}. The maximum theoretical fidelity of $\mathcal{F}_{\rm{max}}=0.9999$ is given by Eq.~(\ref{eq:RosenZener}), and is high due to the small qubit energy splitting, $\delta/(2 \pi) = 150~\rm MHz$, compared to the $\Delta f\approx 19\,\mathrm{GHz}$ bandwidth of the single pulse. To compare, using a hyperfine qubit with 10 GHz splitting, would limit the fidelity to 72\%, demonstrating the benefit of the small splitting of a Zeeman qubit. 

In the experiment, we see additional sources of infidelity. In Fig.~\ref{fig:ultrafast_rabi_single_beam} we see that the visibility scales inversely with the pulse area, or peak Rabi frequency. The additional infidelity can be explained by a thermal spread in the ion's initial position causing different regions of the laser beam to be sampled experiment to experiment. The distribution of Rabi frequencies leads to a decay in the visibility. This type of infidelity in ion traps was studied previously \cite{Cetina_2022}, and an analytic solution for the transition probability can be found \cite{jaffethesis} 
\begin{align}
    P_\downarrow=\frac{1}{2}\left(1-{}_1F_2\left[\frac{g}{2};\frac{1}{2},1+\frac{g}{2};-\theta^2\right]\right),
\end{align}
where $_1F_2[a;b,c;x]$ is a hypergeometric function, $g=\frac{w_0^2}{2\sigma^2_{\rm{ion}}}$ is the ratio of the beam waist to the thermal spread of the ion's position $\sigma_{\rm{ion}}=\sqrt{\frac{k_B T}{m\omega^2}}$ \cite{thermalspread}, with $m$ the ion mass, $T$ the ion temperature, $\omega$ the trap frequency. Fitting the data to this function we extract the needed energy for a single laser pulse to perform a $\pi$ rotation, 38(2)~nJ, as well as the ion temperature, $T=$0.5(1)~mK. Using the state preparation and measurement (SPAM) limits shown in Fig.~\ref{fig:ultrafast_rabi_single_beam}, we applied the Feldman-Cousins method to determine the 90\% confidence interval for the SPAM-corrected fidelity of a $\pi$-pulse, $\mathcal{F}_{\mathrm{thermal}}=(97^{+3}_{-4})\%$. The fidelity central value is given as the ratio of the visibility of the curve, $\mathcal{V}_{\mathrm{fit}}=0.68$, to the SPAM visibility, $\mathcal{V}_{\mathrm{SPAM}}=0.7$, with the visibility taken as the difference between the highest and lowest transition probabilities.

\section{Single-Pulse Spin-Motion Entanglement}

\begin{figure}[!t]
    \centering
    \includegraphics[width=1\linewidth]{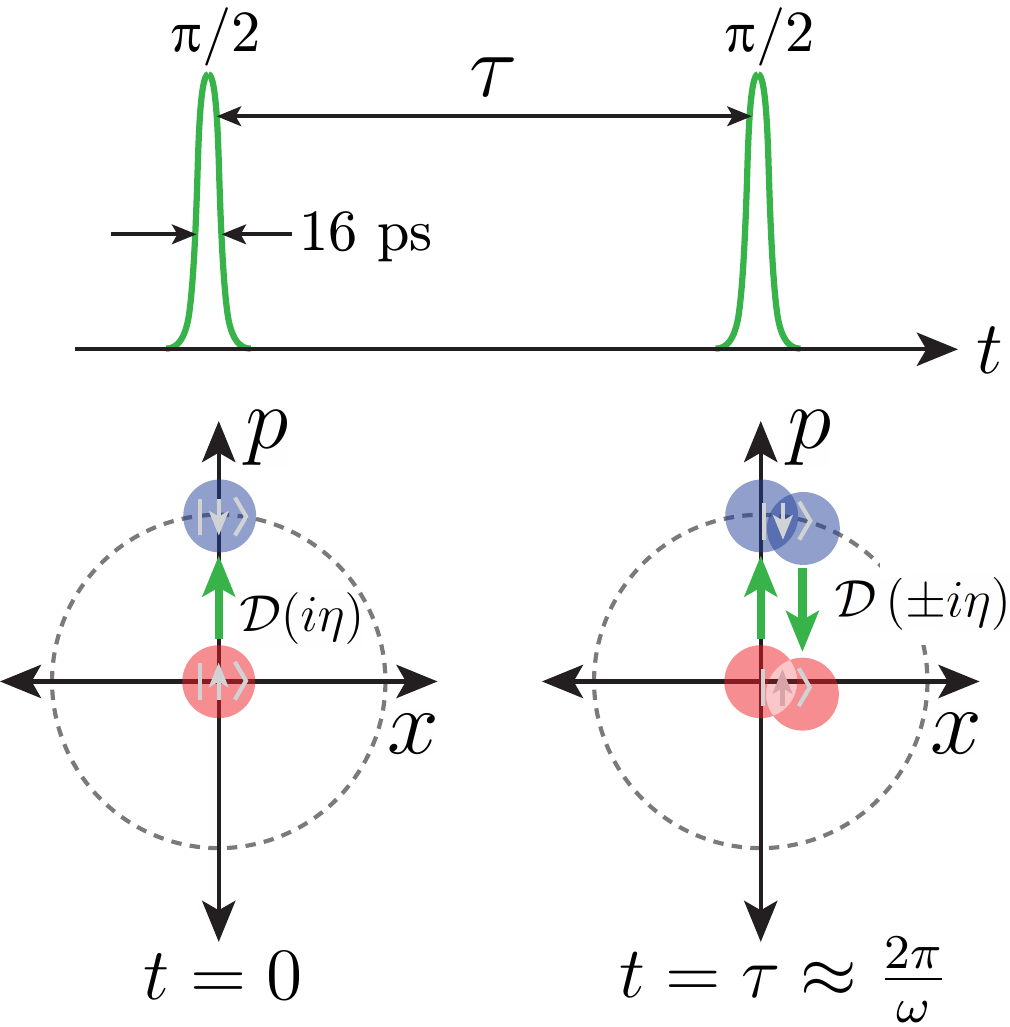}
    \caption{Modulation of wavepacket overlap due to spin-motion entanglement during Ramsey sequence. The evolution of the ion's coherent state is shown in phase space. Left: $\pi/2$ pulse splits wavepacket due to SDK and the kicked wavepacket will oscillate in the harmonic potential. Right: second $\pi/2$ pulse kicks remaining population. Overlap of wavepackets determines fringe visibility.}
    \label{fig:kick_schematic}
\end{figure}
Next, we investigate spin-dependent kicks, a key component for two-qubit gates and matter wave interferometry. We drive stimulated Raman transitions of the qubit in a non-copropagating configuration (Fig.~\ref{fig:ultrafast_paper_exp_figure}(c)), such that the internal state (spin) and motional state are left entangled. The polarization of the light is now correlated with the direction of propagation such that population transfer can only occur by exchanging one photon from each beam, producing a spin-dependent kick: an ion initially in $\ku$ ($\kd$) receives a momentum kick of $\hbar\vec{k}_{\rm eff}$ ($-\hbar\vec{k}_{\rm eff}$) along the trap axial direction, where $\vec{k}_{\rm eff}$ is the wavevector difference between the two beams, $\vec{k}_{\sigma}-\vec{k}_{\pi}$. This is equivalent to a displacement, $\mathcal{D}( i\eta)$, in phase space (Fig.~\ref{fig:kick_schematic}), with $\eta \equiv |\vec{k}_\mathrm{eff}|\sqrt{\frac{\hbar}{2m\omega}}=0.56$ the effective Lamb-Dicke parameter. The kicks are accompanied by a spin flip. Since the pulse width is much shorter than the trap period, we approximate the ultrafast kicks as instaneous momentum displacements and the time evolution operator for an impulsive SDK pulse between degenerate Raman levels becomes:
\begin{align}  
    U(\theta)=\cos\left(\tfrac{\theta}{2}\right)\!\mathds{1}+i\sin\left(\tfrac{\theta}{2}\right)\!\left[\mathcal{D}(i\eta)\hat{\sigma}_-+\mathcal{D}(-i\eta)\hat{\sigma}_+\right],
    \label{eq:evolution}
\end{align}
where $\hat{\sigma}_+$ and $\hat{\sigma}_-$ are the qubit raising and lowering operators, which act on the internal state of the ion and are composed of the usual spin Pauli matrices, $\hat{\sigma}_x$, $\hat{\sigma}_y$, such that $\hat{\sigma}_\pm=\frac{1}{2}\left(\hat{\sigma}_x\pm i\hat{\sigma}_y\right)$. 

To explicitly demonstrate the spin-dependent kick, we prepare the ion in $\ku$ and perform a Ramsey pulse sequence consisting of two SDK $\pi/2$ pulses separated by a variable time, $\tau$, set by the pulse picker, during which the ion undergoes harmonic motion in the trap (Fig.~\ref{fig:kick_schematic}). For each wait time Ramsey interference fringes (see insets of Fig.~\ref{fig:ramsey_revival}) were mapped versus detuning, $\delta$, by scanning the qubit splitting via the static magnetic field. The first SDK $\pi/2$ pulse produces a momentum kick described by a displacement operator $\mathcal{D}(i\eta)$. The time evolution operator during the Ramsey wait time, $U_{\rm wait}$, causes the qubit internal state, $\lvert\psi\rangle$, to pick up a phase $\phi_{\uparrow(\downarrow)}=(-)\delta\tau/2$ and the coherent state evolution, $|\alpha\rangle\rightarrow|\alpha e^{-i\omega\tau}\rangle$, as the ion oscillates in the trap. The degree of wavepacket overlap is encoded in the Ramsey interference fringe visibility $\mathcal{V}$ and is  dependent on the timing of the second $\pi/2$ pulse as illustrated in Fig.~\ref{fig:kick_schematic}. 

The wavepacket overlap after the second $\pi/2$ pulse depends on the initial motional distribution of the ion. Doppler cooling prepares the ion in a thermal state with a mean occupation number $\bar{n}$ in the harmonic oscillator potential. The thermal distribution creates a mixed state, which can be expressed using the Glauber-Sudashan distribution $\rho=\int \diff{\alpha}\  P_G(\alpha)\lvert\psi,\alpha\rangle\langle\psi,\alpha\lvert$, with $P_G(\alpha)=\frac{1}{\pi \bar{n}}e^{-|\alpha|^2/\bar{n}}$ for a thermal distribution of coherent states, $|\alpha\rangle$, with an average occupation number $\bar{n}$. The final density matrix is found by evolving the wavefunction according to Eq.~(\ref{eq:evolution}), $\lvert\psi_\textrm{f},\alpha_\textrm{f}\rangle=U(\tfrac{\pi}{2})U_{\textrm wait}U(\tfrac{\pi}{2})\lvert\psi_\textrm{i},\alpha_\textrm{i}\rangle$. The population of the ion's internal states is then found by tracing out the motion from the density matrix: $P_\uparrow=\langle\uparrow\rvert\textrm{Tr}_\alpha(\rho)\lvert\uparrow\rangle$  \cite{Johnson2015Sensing}. Leading to the probability to remain in the $\lvert\uparrow\rangle$ state, with $\gamma=\delta\tau+\eta^2\sin\omega\tau$:
\begin{align}
    P_\uparrow=\frac12-\frac12\cos\left(\gamma\right)\exp\left[-\eta^2(1-\cos\omega\tau)(2\bar{n}+1)\right].
\end{align}
Scanning over the detuning gives a sinusoidal fringe from which we determine the visibility, $\mathcal{V}$, the difference between the maximum and minimum transition probability:
\begin{align}
    \mathcal{V} = \exp[-\eta^2(1-\cos\omega\tau)(2\bar{n}+1)].
    \label{eq:visib}
\end{align}
Initially, as the wavepackets separate, the visibility rapidly
decays. The wavepackets re-overlap in phase space after an integer number of trap periods, $T=2\pi/\omega$, and the fringe visibility revives as shown in the top plot of Fig.~\ref{fig:ramsey_revival}. The width of the decay and revivals is dependent on the combination of ion parameters $2\eta^2\left(2\bar{n}+1\right)$. 

Fitting the fringe visibility at each value of the wait time gives the data plotted in Fig.~\ref{fig:ramsey_revival}, with error bars given by the fit uncertainty. Representative Ramsey fringes with their fits are inset with arrows indicating the corresponding Ramsey wait time. The red line is a fit to the data according to Eq.~(\ref{eq:visib}) with an additional offset $A =$~0.036(4) that accounts for quantum projection noise during data collection, as well as an overall scaling factor, $B =$~0.41(2) accounting for the finite fidelity of this operation and our readout procedure:
\begin{figure}[!t]
    \centering
    \includegraphics[width=\linewidth]{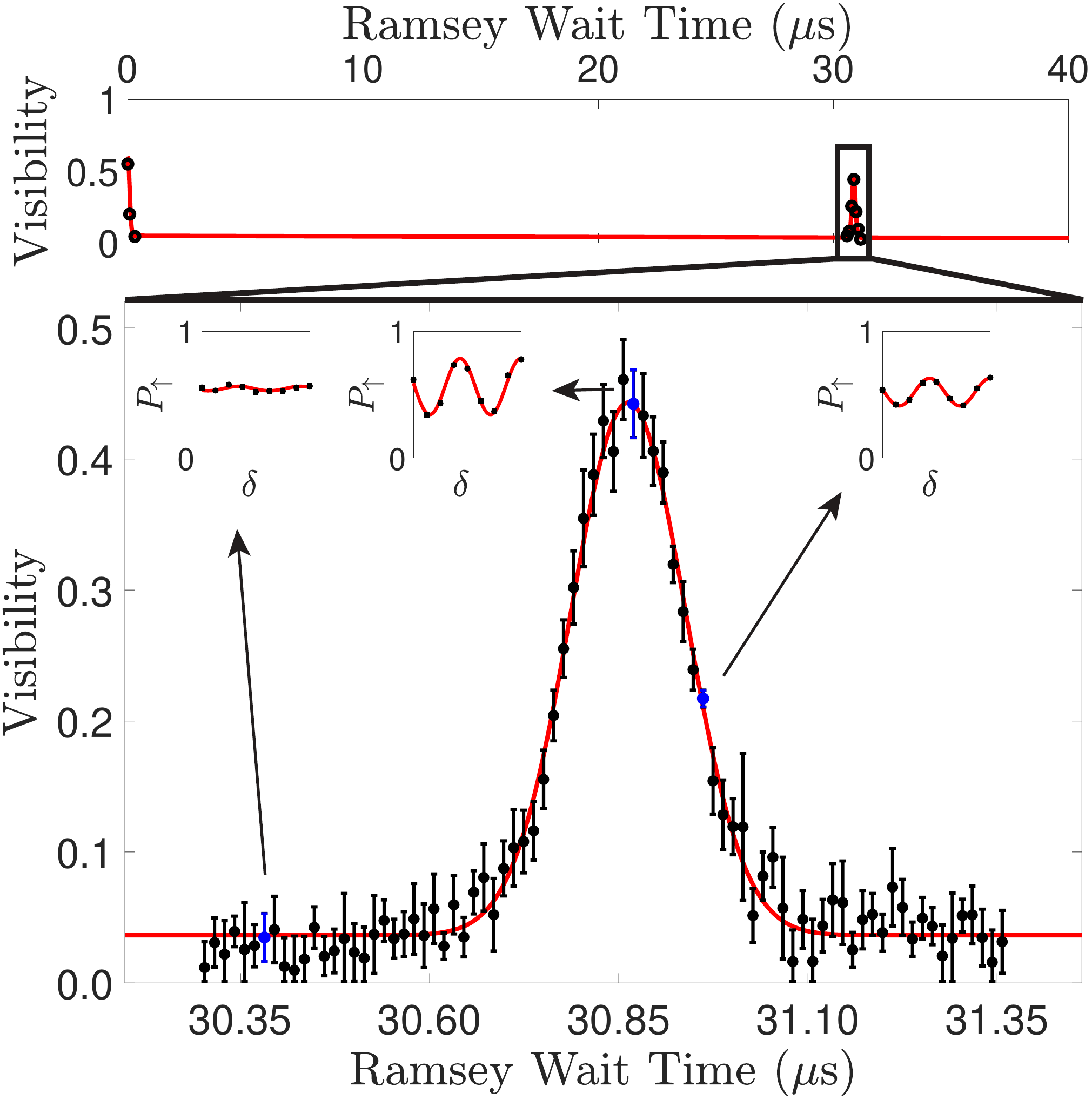}
    \caption{Collapse and revival of spin coherence at the trap period. The topmost plot shows the decay and revival of the Ramsey fringe visibility after the ion has oscillated through one axial trap period. The main plot shows the revival from the top plot and the data points are the best-fit amplitude values of Ramsey fringes as a function of the applied wait time, the red line is a fit to Eq.~(\ref{eq:fitvisibility}). Inset shows some representative fringes.}
    \label{fig:ramsey_revival}
\end{figure}
\begin{align}
    \mathcal{V}&=A+B\exp\left[-\eta^2(1-\cos\omega\tau)(2\bar{n}+1)\right].
    \label{eq:fitvisibility}
\end{align}
From this fit we extract a revival time of $\tau_{\rm rev}=$~30.864(1)~$\upmu$s. The corresponding secular frequency of $\omega=2\pi\times$~32.400(1)~kHz is in good agreement with an independent measurement of the axial secular frequency using the ``tickle scan'' method \cite{Drewsen2004}. The fit value for the mean occupation number $\bar{n}=$~1059(80) implies a temperature of 1.6(1)~mK. 

Using known sources of infidelity, we can account for the max fringe visibility, $\mathcal{V}_{\rm max}=$~0.45(2), of the spin-motion entanglement. The visibility is reduced due to the limited SPAM visibility, the infidelity of the SDK pulses, and the ion coherence time. Unlike in the single-beam single-qubit rotation experiment, the electric field polarization at the ion when using two beams has no $\sigma^+$ component. This leads to a differential light shift of the qubit states dependent on the strength of the $\sigma^-$ polarized beam, as seen in Eq.~(\ref{eq:lightshift}). Lowering the intensity of the $\sigma^-$ polarized beam reduces the differential shift but requires higher intensity of the $\pi$ polarized beam to maintain the same Rabi frequency.  Numerically solving the Schrodinger equation, we find that with a pulse energy of 14~nJ for the $\sigma^-$ polarized beam at about half the energy of the $\pi$-polarized beam (24~nJ) the differential light shift limits our fidelity to $\mathcal{F}_{\rm{lightshift}}=0.95(1)$. We calculate the overall visibility $\mathcal{V}_{\rm tot}=\mathcal{V}_{\rm SPAM}\times\mathcal{F}_{\rm SDK}\times\exp(-\tau_{\rm rev}/T_2)=0.47(4)$, taking $\mathcal{V}_{\rm SPAM}=$ 0.70(3) from Fig.~\ref{fig:ultrafast_rabi_single_beam}, $\mathcal{F}_{\rm SDK}=\mathcal{F}_{\rm{thermal}}\times\mathcal{F}_{\rm{lightshift}}=$~0.92(3), and from auxiliary measurements, the Zeeman qubit coherence time, $T_2\approx 100~\upmu$s.

\section{Conclusion}
\label{sec:conclusions}

We have demonstrated ultrafast control of a trapped ion Zeeman qubit. Using a high-intensity mode-locked laser to drive a Raman transition, we can perform a single qubit $\pi$-rotation using a single laser pulse, in approximately 16.4(5)~ps. The fidelity of this qubit rotation is estimated to be $(97^{+3}_{-4})\%$, currently limited by the ion's thermal position spread. Operating in a non-copropagating geometry, the same Raman transition was used to perform ultrafast spin-motion entanglement, a key ingredient for two-qubit gates and matter-wave interferometry. The momentum imparted by the resulting spin-dependent kick leads to a reduction in the fringe visibility of a Ramsey pulse sequence as the wavepackets separate. Revival of the visibility is observed at a time equal to the trap period, and the variation of the visibility allows us to infer both the efficiency of spin-motion entanglement and the mean occupation number of the trapped ion.

In contrast to previous work with hyperfine qubits \cite{Madsen2006,Campbell2010,Mizrahi2013,Wong-Campos2017}, the fideltiy of the spin-motion entanglement is not limited by multi-photon transitions. Polarization selectivity precludes diffraction of the atomic wavepacket into multiple momentum orders. For applications, such as matter wave interferometry, where higher momentum transfer is beneficial, retroreflecting the SDK beams can double the momentum transfer while returning the ion to its initial state, enabling SDKs to be applied at the repetition rate of the laser. 

Further improvements can be achieved by increasing the fidelity of the state readout, decreasing the ion's thermal position spread, decreasing the differential light shift, and increasing the qubit coherence. Increasing the SPAM fidelity will increase the Ramsey fringe visibility and enable a more precise measurement of the fidelity of ultrafast qubit operations. This can be achieved by using the narrow $^2{\rm S}_{1/2}~{\leftrightarrow}~^2{\rm D}_{5/2}$ transition to perform electron shelving \cite{Dietrich2010,Yum2017}. The ion's spread in position is strongly dependent on the trap secular frequency. For a $\ ^{138}$Ba$^+$ ion cooled to the Doppler limit in a trap with $\omega_{\rm sec}=2\pi\times 200$ kHz this source of infidelity is reduced to the $10^{-5}$ level. By choosing a laser at the ``magic'' wavelength ($\lambda\approx 485\,\text{nm}$ $\mathrm{Ba}^+$, corresponding to frequency $\omega=\omega_0+\omega_{\rm FS}/5$), the differential light shift can be nulled, as seen in Eq.~(\ref{eq:lightshift}). The coherence time can be increased by using permanent magnets to decrease the magnetic field noise. Zeeman qubits coherence times exceeding 1 s were demonstrated in \cite{LongLivedZeeman}. The spin-dependent kick scheme that we have demonstrated here will be harnessed to perform trapped-ion interferometry \cite{Campbell2017,West2019}. \\
\begin{acknowledgments}
This work was supported by the Office of Naval Research (award N000141712256) and the Defense Advanced Research Projects Agency (award D18AP00067).
\end{acknowledgments}
\FloatBarrier

\bibliography{bibl.bib}

\end{document}